\documentclass[a4paper,11pt]{article}
\usepackage{pos}

\newcommand{\mB}{\mathcal{B}}
\newcommand{\mF}{\mathcal{F}}
\newcommand{\mL}{\mathcal{L}}
\newcommand{\mM}{\mathcal{M}}

\newcommand{\mO}{\mathcal{O}}
\newcommand{\mV}{\mathcal{V}}

\title{Discerning EFTs through multi-Higgs production}

\author[a]{Rafael L. Delgado}
\author[b]{Raquel G\'omez-Ambrosio}
\author[c]{Javier Mart\'\i nez-Mart\'\i n}
\author[d]{Alexandre Salas-Bern\'ardez}
\author*[c]{Juan J. Sanz-Cillero}

\affiliation[a]{Dept. Matem\'atica Aplicadas a las TIC, Universidad Polit\'ecnica de Madrid, \\Nikola Tesla, s/n, 28031 Madrid, Spain}
\affiliation[b]{Dipartimento di Fisica, Universit\`a di Torino, and INFN, Sezione di Torino, \\Via P. Giuria 1, 10125 Torino, Italy}
\affiliation[c]{Dept. F\'\i sica Te\'orica and IPARCOS, Universidad Complutense de Madrid, \\Plaza de las Ciencias 1, 28040 Madrid, Spain}
\affiliation[d]{Dept. Análisis Matemático y Matemática Aplicada and IPARCOS, \\Universidad Complutense de Madrid, Plaza de las Ciencias 3, 28040 Madrid, Spain}

\emailAdd{rafael.delgado@upm.es}
\emailAdd{raquel.gomezambrosio@unito.it}
\emailAdd{javmar21@ucm.es}
\emailAdd{alexsala@ucm.es}
\emailAdd{jjsanzcillero@ucm.es}

{ \vbox{\hbox{\hspace*{11.25cm}IPARCOS-UCM-24-054}}}
{ \vbox{\hbox{\hspace*{12.00cm}COMETA-2024-32}}}

\abstract{
In these proceedings we present the main results of~\cite{articulo}, where we explore the phenomenological implications of multi-Higgs boson production through longitudinal vector boson scattering within the framework of Effective Field Theories (EFTs). We derive compact expressions for effective tree-level amplitudes involving up to four final-state Higgs bosons. Subsequently, we compute total cross sections for scenarios relevant to the LHC, where we observe that the general Higgs Effective Theory (HEFT) prediction avoids the strong suppression found in the Standard Model Effective Field Theory (SMEFT), typically expected to be several orders of magnitude smaller.
}

\FullConference{42nd International Conference on High Energy Physics (ICHEP2024)\\
18-24 July 2024\\
Prague, Czech Republic\\}

\begin{document}
\maketitle

\section{Introduction and theoretical framework}

The Standard Model Effective Field Theory (SMEFT) and the Higgs Effective Field Theory (HEFT) provide two distinct frameworks for studying new physics effects beyond the Standard Model (SM) in the Higgs sector.
While SMEFT assumes a linear realization of the Higgs,
HEFT treats the Higgs sector symmetry as a non-linear sigma model, allowing additional interactions between the Higgs and the longitudinal components of the electroweak gauge bosons~\cite{LHCHiggsCrossSectionWorkingGroup:2016ypw}.

In Ref.~\cite{articulo}, we derive compact expressions for tree-level leading-order (LO) amplitudes and cross sections for multi-Higgs production via electroweak vector boson scattering (VBS). Specifically, we compute the processes $W_L^+ W_L^- \to n\times h$ for $n=2, 3, 4$.
Although during the last years we have witnessed a great improvement in the determination of the $h\to WW$ and $hh\to WW$ couplings from one and two Higgs production analyses~\cite{CMS:2024awa}, we are aware of the difficulties of these experimental studies. Nonetheless, although processes with three or more final-state Higgs bosons are complicated—or simply impossible—to produce at current colliders, next runs and future facilities~\cite{Brigljevic:2024vuv}
are expected to be able to constrain
couplings that are nowadays out of reach.

We focus our analysis on the vector boson scattering $W^+_LW^-_L\to n\times h$ in the kinematic region well over the production threshold ($s\gg m_W^2$). Hence, we will neglect the electroweak (EW) gauge and Higgs boson masses (as $m_h\sim m_{W,Z}$).
Our HEFT computation will extract the amplitudes at leading order (LO) in the chiral expansion, $\mathcal{O}(p^2)$, so we will need the EFT operators with two derivatives. Mass terms and operators with a higher number of derivatives are not considered.
Under the referred kinematic approximation $s\gg m_{W,Z}^2$ (where phenomenologically $m_{W,Z}^2 \sim m_h^2$), we will use the Equivalence Theorem (EqTh)~\cite{Veltman:1989ud}
and approximate the scattering of longitudinal EW gauge bosons with the scattering of Goldstone bosons $\omega^a$,
$     \mM(W_LW_L\to n\times h) \,\simeq \, -\, \mM(\omega\omega\to n\times h)$, up to corrections which are suppressed by the gauge boson masses $m_{W,Z}$ over  $\sqrt{s}$.

For this work~\cite{articulo}, we have developed specific tools that are publicly available: a specific stand-alone \textsc{Mathematica} code for the generation of arbitrary $\omega\omega\to n\times h$
amplitudes~\footnote{ \href{https://github.com/alexandresalasb/WWtonHcalculator}{https://github.com/alexandresalasb/WWtonHcalculator} .  };
we have constructed a HEFT model file\footnote{
\href{https://github.com/Javomar99/EWET}{https://github.com/Javomar99/EWET} ,
\href{https://github.com/Javomar99/Multi_Higgs_HEFT}{https://github.com/Javomar99/Multi\_Higgs\_HEFT}  \cite{Martinez-Martin}.}
in \textsc{FeynRules}~\cite{Rules} and \textsc{FeynCalc}~\cite{Calc} for the analytical calculations of the VBS amplitudes.
Also, the HEFT model file was used in combination with \textsc{MadGraph5}$\_$\textsc{aMC}~\cite{Alwall:2011uj} for a numerical test of the analytical expressions.
Finally, as the $m_{W,Z,h}=0$ limit is considered here, we developed a simple and efficient multiparticle phase-space numerical integration code for massless final states, {\tt MaMuPaXS}~\footnote{
\href{https://github.com/mamupaxs/mamupaxs}{https://github.com/mamupaxs/mamupaxs} .
}.

We will work at the lowest order approximation in the equivalence theorem and neglect mass corrections. Likewise, we will perform our effective theory calculation at leading order (LO), $\mO(p^2)$, so we will just consider the part of LO Lagrangian~\cite{LHCHiggsCrossSectionWorkingGroup:2016ypw},
\begin{equation}\label{eq:LOlag}
    \mL_{\rm HEFT}\, =\, \frac{1}{2}(\partial_\mu h)^2 + \frac{v^2}{4}\mF(h)\, \mbox{Tr} \left\{\partial_\mu U^\dagger \partial^\mu U\right\}\, ,
\end{equation}
with the EW vacuum expectation value $v\simeq 246$~GeV. Here we are only showing the relevant $\mO(p^2)$ derivative operators that may contribute to the HEFT $\omega\omega\to n\times h$ amplitude at LO, provided by tree-level diagrams with the $\mO(p^2)$ derivative vertices.
As the Higgs field is an EW singlet in the HEFT approach, the non-linear sigma model term containing the EW Goldstones through $U(\omega)= 1+ i\sigma^a \omega^a+\mO(\omega^2)$ can be multiplied by any arbitrary flare function, $\mF(h)$, and still remain symmetry invariant. In general, this function will have an expansion in the Higgs field in the form,
\begin{equation}
    \mF(h)\, = \,1 \, +\, a_1 \, \left(\frac{h}{v}\right) \, +\, a_2 \, \left(\frac{h}{v}\right)^2 \, +\, a_3 \, \left(\frac{h}{v}\right)^3 \, +\, a_4 \, \left(\frac{h}{v}\right)^4 \, +\, ...
\end{equation}
where the effective coupling constants $a_n$ parametrize the $\omega \omega\to n\times h $ Feynman rules and are not fixed by symmetry. Their values will need to be fixed by matching with a high-energy underlying theory (top-down approach) or from experiment (eventually providing constraints on possible SM extensions in bottom-up approaches). Although the labelling $a_n$ provides a systematic naming for these interactions, the one and two Higgs couplings are often referred as $
    a\,\equiv\, \kappa_V\,\equiv\, a_1/2\, $ and $\;
    b\,\equiv \, \kappa_{2V}\, \equiv \, a_2$,
where one recovers the SM flare function $\mF(h)=(1+h/v)^2$ in the limit $a,b\to 1$.

Regarding the $\omega\omega\to n\times h$ couplings, one has  less freedom in SMEFT. Up to next-to-next-to-leading order in the effective theory expansion, there are only two free couplings that preserve custodial symmetry entering the Lagrangian in (\ref{eq:LOlag}): $c_{H\Box}^{(6)}$ at $\mO(\Lambda^{-2})$ and  $c_{H\Box}^{(8)}$ at $\mO(\Lambda^{-4})$. Thus, the contributions from the derivative dimension--6 and dimension--8 SMEFT operators introduce deviations in the SM flare function, with the $a_j$  provided up to $\mO(\Lambda^{-4})$ by~\cite{Gomez-Ambrosio:2022qsi}:
\begin{align}
&a_1/2\,= \, a \, = \, 1
\, +\,  \frac{d}{2}\, +\, \frac{d^2}{2}\left(\frac{3}{4} +\rho\right)      \,+\,\mathcal{O}\left(d^3\right)\, ,
&a_2 \,= \, b \, = \, 1
\, +\,  2d  \, +\, 3 d^2\left(1+\rho\right) \,+\,\mathcal{O}\left(d^3\right)\, ,
\nonumber\\
&a_3 \,=  \frac{4}{3}d  \, +\,  d^2\left(\frac{14}{3}+4\rho\right) \,+\,\mathcal{O}\left(d^3\right)\, ,
&a_4 \,=  \frac{1}{3}d  \, +\,  d^2\left(\frac{11}{3}+3\rho\right) \,+\,\mathcal{O}\left(d^3\right)\, ,
\nonumber\\
&a_5 \,=   d^2\left(\frac{22}{15}+\frac{6}{5}\rho\right) \,+\,\mathcal{O}\left(d^3\right)\, ,
&a_6 \, = d^2\left(\frac{11}{45}+ \frac{1}{5}\rho\right) \,+\,\mathcal{O}\left(d^3\right)\, ,
\label{eq:SMEFT-FF}
\end{align}
with
$d=\frac{2v^2 c_{H\Box}^{(6)}}{\Lambda^2}$
and $\rho=\frac{c_{H\Box}^{(8)}}{2 (c_{H\Box}^{(6)})^2}\;
$, being $c_{H\Box}^{(D)}$ the dimensionless Wilson coefficient for the operator $|H|^{D-4} \Box |H|^2/\Lambda^{D-4}$. Higher coefficients $a_n$ with $n\geq 7$ vanish at this order in SMEFT.

In this sense, one observes the SM is a particular limit of SMEFT, and SMEFT itself is a particular case of HEFT: the $n\times h$ couplings $a_1,\, ... \, a_6$ are correlated and determined by a much smaller amount of SMEFT parameters, namely $c_{H\Box}^{(6)}$ and $c_{H\Box}^{(8)}$~\cite{Gomez-Ambrosio:2022qsi}.

\section{Multi-Higgs Production from VBS}

Multi-Higgs production through VBS is particularly interesting as it probes the high-energy behavior of the electroweak sector. At high energies, the longitudinal components of the gauge bosons ($W_L, Z_L$) dominate the scattering processes. In the SM, the scattering amplitudes for processes like $W_L W_L \to hh$ are suppressed by the Higgs self-coupling, which is small. However, in HEFT, new interaction terms between the Higgs and the Goldstone bosons lead to enhanced amplitudes.
In this work, we compute tree-level amplitudes for processes involving up to four Higgs bosons in the final state. In general we will be providing here the results for the amplitudes with $\omega^+\omega^-$ initial state~\cite{articulo}:
\begin{equation}
T_{\omega\omega\to 2h}
\,=\, -\, \frac{\hat{a}_2   s}{v^2} \,  ,
\qquad
    T_{\omega\omega\to 3h}
    \,=\,
    \,-\, \frac{3 \hat{a}_3 s}{v^3} \,,
\qquad
    T_{\omega\omega\to 4h} \,=\,
  \, -\, \frac{4 s}{v^4}\left(3\hat{a}_4 + \hat{a}_2^2 \, (B-1)  \right)        \, ,
\label{eq:Tnh-HEFT}
\end{equation}
with $\hat{a}_2=a_2-a_1^2/4 = b-a^2$, $\hat{a}_3 = a_3- \frac{2}{3}a_1 \left(a_2 -  a_1^2/4\right) = a_3 - \frac{4}{3} a \left(b-a^2\right)$ and .
$\hat{a}_4 =  a_4 -   \frac{3}{4} a_1 a_3 + \frac{5}{12}a_1^2\left(a_2-a_1^2/4\right) = a_4 - \frac{3}{2} a\,a_3 + \frac{5}{3}a^2\left(b-a^2\right)$.
The dimensionless function $B(k_i; p_j)$ is a combination provided by one-crossed-channel-propagator topologies
~\cite{articulo}.
\footnote{
$ B= f_1f_2f_3f_4 \,\bigg(\mathcal{B}_{1234} + \mathcal{B}_{1324} + \mathcal{B}_{1423} + \mathcal{B}_{2314} + \mathcal{B}_{2413} + \mathcal{B}_{3412}\bigg)$,
with each term $\mathcal{B}_{ijk\ell} \,=\, \frac{z_{ij}z_{k\ell}}{2f_if_jz_{ij}-f_iz_i-f_jz_j}$ only depending on external momenta scalar products of the form $f_i = q p_i/q^2$, $z_i = 2 k_1 p_i/q p_i$, $z_{ij} = z_{ji}= q^2\, (p_ip_j)/[(q p_i)\, (q p_j)]$. The total four-momentum is denoted as $q=k_1+k_2=p_1+p_2+p_3+p_4$ and the on-shell condition $k_{1,2}^2=p_{1,2,3,4}^2=0$.
Note that here we have rewritten the expression $f_1f_2f_3f_4 \mB_{ijk\ell}=(p_i+p_j)^2 (p_k+p_\ell)^2/[4 q^2 (k_1-p_i-p_j)^2]$ for on-shell massless external particles in a more convenient form for the phase-space integration.
In the center-of-mass (CM) rest-frame these relations define:
the three-momentum fractions $f_i= \lVert\vec{p}_i\rVert/\sqrt{s}$ ($s=4\lVert\vec{k}_1\rVert^2$) for each outgoing Higgs boson;
the angular functions $z_i= 2\sin^2(\theta_i/2)$ with $\theta_i$ being the angle between the $i$-th Higgs boson and the incoming $\omega^+$
Goldstone boson momenta, $\vec{k}_1$
(that is, $z_1=1-\cos\theta$, $z_2=1+\cos\theta$ as usual in a two-body problem with $t$ and $u$ channels);
$z_{ij}= 2\sin^2(\theta_{ij}/2)$,  with $\theta_{ij}$ being the angle between the $i$-th and $j$-th Higgs bosons.
We remark that by construction we have factored out the scaling with $s$ so that the kinematic function $B$ only depend on the angular distribution of the Higgses.}

The phase-space integration can be readily performed for $\omega^+(k_1)\, \omega^-(k_2) \to h(p_1)\, h(p_2) $ and $\omega^+(k_1)\, \omega^-(k_2) \to h(p_1)\, h(p_2)\, h(p_3) $, while for $\omega^+(k_1)\, \omega^-(k_2) \to h(p_1)\, h(p_2)\, h(p_3)\, h(p_4)$ we needed to develop a numerical integration code~\cite{articulo}:
\begin{eqnarray}
& \sigma_{\omega\omega\to 2h}
\,=\,  \frac{8\pi^3 \, \hat{a}_2^2 }{s}\, \left( \frac{s}{16\pi^2 v^2}\right)^2\, ,
\qquad
\sigma_{\omega\omega\to 3h} \,=\,
\frac{12\pi^3 \,\hat{a}_3^2} {s}\left(\frac{s}{16\pi^2 v^2}\right)^3\, , &
\nonumber
\\
&\sigma_{\omega\omega\to 4h} \,=\,   \frac{8\pi^3}{9 s}   \left(\frac{s}{16\pi^2 v^2}\right)^4
\left[\left(3\hat{a}_4 - \hat{a}_2^2 \right)^2   + 2 \left(3\hat{a}_4 - \hat{a}_2^2 \right) \hat{a}_2^2 \chi_1 +\hat{a}_2^4\chi_2\right] \, ,&
\end{eqnarray}
with $\chi_n =\mathcal{V}_4^{-1}\int d\Pi_4 \,  B^n$ and $\mV_4= s^2 / \left[24 (4\pi)^5 \right]$.
evaluated them numerically through our phase-space integration code ({\tt MaMuPaXS})~\cite{articulo}:
\begin{eqnarray}
\chi_1 =
 -0.124984\, (10)\,, \qquad
\chi_2 =
0.0193760 \, (16)\, .
\label{eq:chi-n_num}
\end{eqnarray}

One can observe that the $2h$ and $3h$ amplitudes in~(\ref{eq:Tnh-HEFT}) are pure $J=0$, $s$--wave. They only depend on the total CM energy, not on the scattering angles. This is roughly the angular distribution observed in analyses beyond the equivalence theorem for $WW\to 2h$~\cite{Davila:2023fkk}  --up to mass corrections in the forward and backward directions--, very different to SM prediction.
The $4h$ production case is a bit more complicated: the kinematic function $B$ introduces a non-trivial angular distribution for the differential $\omega\omega\to 4h$ cross section, which is no longer a pure $s$--wave. However, we observe a numerical suppression $|\chi_{2}|\ll |\chi_1|\ll 1$. Hence, in a first approximation, one finds a $J=0$ wave numerical dominance.
These distinct angular s-wave distributions should serve as an important discriminating factor in current and future experimental analyses for the VBS production of $2h$~\cite{CMS:2024awa}, $3h$~\cite{Brigljevic:2024vuv}, and even $4h$ final states in some future.

This result is in agreement with previous effective theory $WW\to 2h$ studies~\cite{Anisha:2022ctm}
at high energies, where the Equivalence Theorem approximation is applicable.
Some other works that have analyzed the $WW\to 3h$ scattering in the context of effective theories can be found in~\cite{Gonzalez-Lopez:2020lpd}.

\section{Amplitude suppression in SMEFT}

\begin{figure}[t!]
\center
  \includegraphics[width=0.5\textwidth]{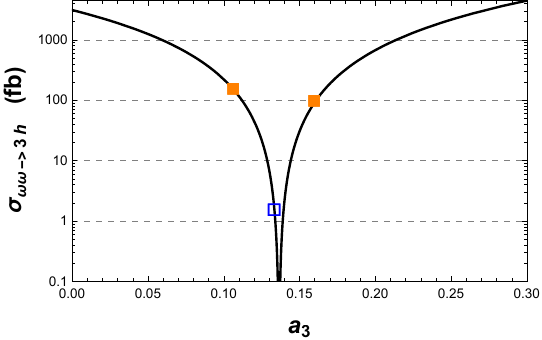}
    \caption{{\small Scan of the $\omega\omega\to 3h$ cross section predictions in terms of $a_3$ at $\sqrt{s}=1$~TeV. The inputs $a_1=a_1^{\rm SMEFT (D=6)}=2.1$ and $a_2=a_2^{\rm SMEFT (D=6)}=1.2$ correspond to the SMEFT$^{\rm (D=6)}$ BP. We have marked a few especial points:  $a_3=a_3^{\rm SMEFT (D=6)}=0.1\overline{3}$        
    (empty blue square) and their 20\% deviations (full orange squares), $a_3=80\% \times a_3^{\rm SMEFT (D=6)}$ and  $a_3=120\% \times a_3^{\rm SMEFT (D=6)}$.
    }}
    \label{fig:scan-pheno-3h}
\end{figure}

We have observed that the $\omega\omega\to n\times h$ amplitude showed a very simple dependence on the HEFT parameters: $\omega\omega\to 2 h$ only depends on the combination $\hat{a}_2$;  $\omega\omega\to 3 h$ only depends on the combination $\hat{a}_3$;  $\omega\omega\to 4 h$ only depends on the effective combinations $\hat{a}_4$ and $\hat{a}_2$.
From~(\ref{eq:SMEFT-FF}) we can read these $a_j$ combinations in SMEFT:
\begin{eqnarray}
\hat{a}_2 = d\, +\, 2 d^2 (1+\rho)\, +\, \mO(d^3)\, ,\quad
\hat{a}_3 =
\frac{4}{3} d^2 (1+\rho)\, +\, \mO(d^3)\, , \quad
\hat{a}_4  =
\frac{1}{3} d^2  (  2
+\rho)\, +\, \mO(d^3)\, ,
\label{eq:SMEFT-ajhat}
\end{eqnarray}
with,
$d=\frac{2v^2 c_{H\Box}^{(6)}}{\Lambda^2}$
and $\rho=\frac{c_{H\Box}^{(8)}}{2 (c_{H\Box}^{(6)})^2}\;,
$. Note that all the $\hat{a}_2$, $\hat{a}_3$, $\hat{a}_4$... vanish in the SM limit.

Thus, we explicitly observe that SMEFT introduces a larger $1/\Lambda^m$ cross-section suppression for an increasing number of final Higgs bosons: $\sigma_{\omega\omega\to 2h}\sim s/\Lambda^4$, and $\sigma_{\omega\omega\to 3h, \, 4h} \sim s^3/\Lambda^8$. The next possible processes follow this trend ($\sigma_{\omega\omega\to 5h, \, 6h} \sim s^5/\Lambda^12$, $\sigma_{\omega\omega\to 7h, \, 8h} \sim s^7/\Lambda^16$, etc.), as proved in~\cite{articulo}.

This high suppression pattern is not due to the size of the deviations $\Delta a_j=a_j-a_j^{\rm SM}$ from the SM couplings $a_1^{\rm SM}=2$, $a_2^{\rm SM}=1$, $a_{k\geq 3}^{\rm SM}=0$.
In principle for small $\Delta a_j\sim v^2/\Lambda^2$ the pure-HEFT amplitudes $T_{\omega\omega\to n\times h}$ will be suppressed by $\Delta \hat{a}_j\sim v^2/\Lambda^2$. However, SMEFT-type theories have very important correlations and cancellations between the $a_j$. Thus, although in general the relevant $\hat{a}_\ell$ combinations for $\omega\omega\to n\times h$ would be expected to be of the order $\hat{a}_\ell\sim \Delta a_j \sim v^2/\lambda^2$, they are fine-tuned in SMEFT to be suppressed by more and more powers of $(v^2/\Lambda^2)$ as one has more and more Higgs bosons in the final state.

This SMEFT fine tuning will be illustrated here with the $\omega\omega\to 3h$ cross section.   We will be using the $D=6$ coupling values
at the benchmark point (BP) with $d=0.1$~\cite{articulo}: $
a = \frac{a_1}{2}
=1.05\, $, $
b= a_2
=1.20\,,\;
a_3
= 0.1\overline{3}                    
a_4
= 0.0\overline{3}$,                  
with $a_{k\geq 5}=0$ for all the remaining couplings.
This SMEFT prediction is then compared in that plot with a pure-HEFT scenarios.
In particular one might think that a ``mild'' modification such as setting $a_3$ to its SM value $a_3^{\rm SM}=0$, while leaving all the other $a_{k\neq 3}$ unchanged, should introduce an important variation of the cross section.
However this is not true: SMEFT correlations between $a_k$ must be very precisely tuned. A small independent deviation in just one of them can lead to a variation of the cross section by several orders of magnitude.

Fig.~\ref{fig:scan-pheno-3h} presents a scan of the $\omega\omega \to 3h$ cross-section at $\sqrt{s}=1$~TeV in terms of $a_3$, analogous to that performed in Ref.~\cite{Englert:2023uug} for $\omega\omega \to 2h$ with $a_2$.
We have plotted the SMEFT$^{\rm (D=6)}$ BP (empty blue square).
one can easily observe how fine-tuned this value is: for $a_1$ and $a_2$ fixed to the BP values, a $\pm20\%$ variation in $a_3$ w.r.t. the  SMEFT$^{\rm (D=6)}$ value leads to an increase in the cross section of orders of magnitude.
Taking $a_3$ to its SM value $a_3^{\rm SM}=0$ does not improve the situation; it actually makes it worse, increasing the cross section by roughly three orders of magnitude.

Due to the general scaling of the amplitude with $s$ in the regime where our EqTh approximation applies, this result holds generally for any energy range within the validity of the EFT (either SMEFT or HEFT). Indeed, one cannot extend our cross section predictions for arbitrary high energies as eventually they exceed the unitarity bound~\cite{Lang:2021hnd}.

\section{Conclusions}

We have shown that the $\omega\omega \to 2h,\, 3h,\, 4h$ cross sections at LO show a very simple structure in the kinematic region of the EqTh approximation considered in this work~\cite{articulo}. Moreover, very specific combinations of couplings, $\hat{a}_2$, $\hat{a}_3$ and $\hat{a}_4$, rule these cross sections. Other dependencies appear only as higher order effects, either in the chiral expansion in powers of momenta or masses, or corrections to the naive EqTh approximation.

We conclude that, as expected, from general arguments the SMEFT cross sections are suppressed by several orders of magnitude with respect to those of non-SMEFT theories.
Our outcomes show that pure-HEFT scenarios predict significantly larger cross sections for processes involving three or more Higgs bosons compared to SMEFT. This enhancement provides a potential signature for new physics beyond the SM, to 
be probed in future LHC runs or at next-generation colliders.

\section*{Acknowledgements}

JJSC wants to thank the ICHEP 2024 organizers for the nice scientific environment.
We would also like to thank M. Cepeda, F.J. Llanes Estrada, E. Mart\'\i n Viscasillas, I. Rosell and D. Zeppenfeld for useful comments and discussions.
This work has been supported by grants PID2022-137003NB-I00, PID2023-148162NB-C21 and PID2021-124473NB-I00 of the Spanish MCIN/AEI/10.13039/501100011033/; EU’s Next Generation grant DataSMEFT23 (PNRR - DM 247 08/2); EU’s 824093 (STRONG2020); EU's COMETA COST Action CA22130; and Universidad Complutense de Madrid under research group 910309 and the IPARCOS institute.

\end{document}